\documentclass[twocolumn,showpacs,superscriptaddress,prl]{revtex4}
\usepackage{psfrag,amsbsy,amssymb,amsmath,bm,graphicx}

\usepackage{color}
\newcommand{\ignore}[1]{}
\newcommand{\bComment}[1]{}
\newcommand{\yComment}[1]{}
\newcommand{\ket}[1]{|#1\rangle}

 \renewcommand{\bComment}[1]{\textcolor{blue}{Boris: #1}}
 \renewcommand{\yComment}[1]{\textcolor{green}{Yuri: #1}}

\begin{document}
\title{Models of environment and $T_1$ relaxation in Josephson Charge Qubits}
\author{Lara Faoro}
\affiliation{Department of Physics and Astronomy, Rutgers University, 
136 Frelinghuysen Road, Piscataway 08854, New Jersey, USA}
\author{Joakim Bergli}
\email{jbergli@fys.uio.no}
\affiliation{Physics Department, 
Princeton University, Princeton, New Jersey 08544, USA}
\author{Boris L. Altshuler}
\affiliation{Physics Department, 
Princeton University, Princeton, New Jersey 08544, USA}
\affiliation{NEC Laboratories America, Inc., 4 Independence Way, 
Princeton, NJ 08540,USA}
\author{Y. M. Galperin}
\affiliation{Department of Physics, University of Oslo, PO Box 1048
  Blindern, 0316 Oslo, Norway}
\affiliation{A. F. Ioffe  Physico-Technical Institute of Russian Academy of
Sciences, 194021 St. Petersburg, Russia}
\affiliation{Argonne National Laboratory, 9700 S. Cass av., IL 60439, USA}

\date{ 
\today }

\begin{abstract}

A theoretical interpretation of the recent experiments of Astafiev 
{\em et. al.} on the $T_1$-relaxation rate in Josephson Charge Qubits is 
proposed. The experimentally observed reproducible nonmonotonic dependence of
$T_1$ on the splitting $E_J$ of the qubit levels suggests further specification
of the previously proposed models of the background charge 
noise.
From our 
point of view the most promising is the ``Andreev fluctuator'' model of 
the noise. In this model the fluctuator is a Cooper pair that tunnels 
from a superconductor and occupies a pair of localized electronic
states. Within this model one can naturally explain both the average linear
$T_1(E_J)$ dependence and the irregular fluctuations. 

\end{abstract}

\pacs{}

\maketitle

Proposals to implement qubits using superconducting nanocircuits have
undergone an amazing development during the last years
\cite{NAKA99,VION02,YU,MARTI02,CHIORES}.  In Josephson Charge Qubit
(JCQ) information is encoded in the charge states of a Cooper pair box.
The JCQ is manipulated 
by tuning gate voltage and
magnetic flux.  Both time resolved coherent oscillations in single and
coupled JCQ have been recently observed
\cite{NAKA99,PASHK03}.  Although decoherence is a severe limitation to
the performances of these devices the dominant source of noise
is yet to be identified.

A significant step towards a characterization of the
environment in a JCQ has
been recently made by Astafiev
{\em et.al.} \cite{astafiev}
The experimental set-up consists of a Cooper pair box connected to a reservoir
through a tunnel junction of SQUID geometry with Josephson energy $E_J$
pierced by an external magnetic field. 
Provided that $E_c\gg E_J\gg T$ (where $E_c$, $E_J$ and $T$ are correspondingly
charging energy, Josephson energy and temperature, $k_B=\hbar=1$).
only two charge states
$\ket{0}$ and $\ket{1}$ are relevant and the Hamiltonian of the box reads:
\begin{equation}
 H_q = -\frac{\delta E_c}{2}\sigma_z -\frac{E_J}{2}\sigma_x \;
\label{Cbox}
\end{equation}
where $\displaystyle{\delta E_c=E_c (1- C_g V_g/e)}$, $C_g$ is the
gate capacitance, $V_g$ is the gate voltage and $e$ denotes the
electron charge.
In the rotated basis  $\{\ket{+}$, $\ket{-}\}$  
the Hamiltonian (\ref{Cbox}) reads:
\begin{equation}
 H_q = -\frac{E}{2}\rho_z,  \qquad \rho_z=\sigma_z  \cos\theta + \sigma_x \sin\theta \;
\label{hamrot}
\end{equation}
where
$\displaystyle{E = \sqrt{\delta E_c^2+E_J^2}}$ and $\theta=
\arctan(E_J/\delta E_c)$. 
One can distinguish the off degeneracy working points 
($\theta\approx0$ and $\delta E_c \gg E_J$) and the
degeneracy one ($\theta=\pi/2$ and  $\delta E_c=0$).
Astafiev {\em et.al.} \cite{astafiev} measured 
the energy relaxation rate $\Gamma_1$
of the JCQ in a wide range of parameters.
Two main features have been observed: (*) Linear increase of  $\Gamma_1$ with
 $E_J$ at large  $E_J$, and 
 (**) Small non-monotonous fluctuations in the $\Gamma_1(E_J)$-function on this
linear background. 

We do not believe that the existing experimental information is sufficient
to identify a unique interpretation. However, it substantially reduces the 
range of possibilities.
In this Letter we show that some models which
have been used to
study dephasing in JCQ can not explain these
features. We propose a model where all of them 
appear naturally.

 Many different
mechanisms can be responsible for decoherence in JCQ.   We
will consider three models, all based on the idea that the oxide layer
close to some metallic reservoir, like one of the leads or gates or 
Cooper pair box itself, 
is disordered and thus hosts trapping
centers, i.e. localized states for the electrons. 
\begin{figure}[h]
\includegraphics[width=8cm]{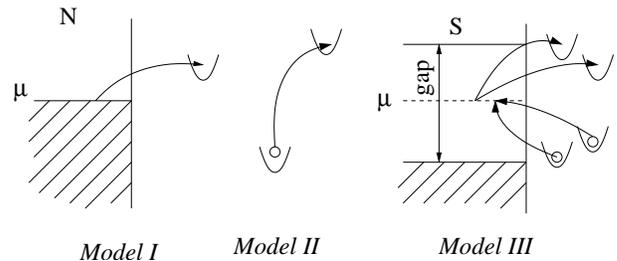}
\caption{
\label{fig0}The three models}
\end{figure}
{\em Model I}: The
reservoir is a normal metal, and electrons can tunnel 
from any state below the chemical potential to an unoccupied trap above the
chemical potential or from an occupied trap to an extended state above the
chemical potential.  {\em Model II}: There is no tunneling between the
reservoir and the traps, but an electron in an occupied trap below the
chemical potential can be excited into an empty trap.  {\em Model III}: The 
reservoir is a superconductor and
a Cooper pair can split in such a way that
each of the electrons end up at an empty trap (Andreev fluctuator).  
At the end of the paper we  
discuss why we do not believe that dephasing through phonon or photon modes 
can describe the experimental results.

To describe the qubit interacting with the environment one
has to supplement Eq. (\ref{Cbox}) by the Hamiltonians of the
environment, $H_E$, the interaction $H_I$, and the tunneling $H_T$. 
Regardless to its particular features an environment is coupled to a JCQ 
through the charge degree of freedom:
$\displaystyle{\sigma_z=\rho_z  \cos\theta - \rho_x \sin\theta}$.
Since only the term $\rho_x$ in the interaction
Hamiltonian will change the state of the qubit, the
relaxation rate $\Gamma_1$ should be proportional to $\sin^2\theta$
if  the qubit-environment 
coupling is weak. 

The charge of the qubit affects the environment in two ways. First, the 
electrostatic interaction shifts the energy of the localized states. Second,
the amplitude $t$ of the tunneling between the trap and the reservoir
(another trap) depends on the qubits charge:
$\displaystyle{t(\sigma_z) = t_0 + \tilde{t} \sigma_z}$.

The total Hamiltonian can be written as
\begin{equation}
H = H_q + H_I + H_t + H_E.
\end{equation}

For the environment Hamiltonian we write 
\[
\begin{split}
 H_E^{(I)} &= \sum_\alpha \epsilon_\alpha c^\dag_\alpha c_\alpha
      + \sum_k\Omega_k f_k^\dag f_k,  \\
 H_E^{(II,III)} &= \sum_\alpha \epsilon_\alpha c^\dag_\alpha c_\alpha.
\end{split}
\]
where  $c_\alpha,(c^\dagger_\alpha)$
destroys (creates) an electron in trap $\alpha$ and $f_{k}, (f^{\dagger}_{k})$ destroys (creates) an electron in the reservoir.
(We assume that the superconductor in {\em Model III} is 
always
in the ground state). 

$H_t$ describes the tunneling in the absence of the qubit
\[
\begin{split}
  H_t^{(I)} &= 
     t_0 \sum_{k,\alpha}(c^\dag_\alpha f_k + c_\alpha f^{\dagger}_{k}), \\
  H_t^{(II)} &= 
     t_0 \sum_{\alpha\neq\beta}c^\dag_\alpha c_\beta, \\
  H_t^{(III)} &= 
     t_0 \sum_{\alpha\neq\beta}(c^\dag_\alpha c^\dag_\beta + c_\alpha c_\beta).
\end{split}
\]

The coupling of 
the qubit with the environment is governed by the Hamiltonian
\begin{equation}\label{HI}
  H_I = 
   \left(v\sum_\alpha c^\dag_\alpha c_\alpha + \frac{\tilde{t}}{t_0}H_t\right)
\sigma_z
\end{equation}

Let us now discuss processes where the qubit, initially prepared in the
excited state, releases the energy by exciting the environment 
($T_1$-processes).
When the coupling
is weak
the relaxation rate can be 
derived by using the Fermi Golden Rule.
If originally the qubit is in the excited state and the bath occupies a state
$|i\rangle$ with probability $\rho_i$ the
decay rate to the ground state, $\Gamma_{\downarrow}$, is 
\begin{eqnarray}
\Gamma_{\downarrow}&=& 2 \pi \sum_{i,f} \rho_i^0 | \langle +,f|T e^{-i \int_0^t H_I(t') dt'}|-,i \rangle |^2 \;.
\label{FermiGR}
\end{eqnarray}
Here $|f\rangle $ is the final state of the
bath.
If the qubit is prepared in the ground state the transition rate to the
excited state is 
at thermal equilibrium $\displaystyle{
\Gamma_{\uparrow}= e^{-E/T } \Gamma_\downarrow}$.
The total relaxation rate is then  
$\displaystyle{\Gamma_1 = \Gamma_{\downarrow}+ \Gamma_{\uparrow}
=(1+e^{-E/T})\Gamma_{\downarrow}}$. In the limit
$\displaystyle{E\gg T}$ we have
$\displaystyle{\Gamma_1\approx\Gamma_\downarrow}$.

When using Eq. (\ref{FermiGR}) the exponential has to be expanded to
the necessary order. Notice that the second term in Eq. (\ref{HI}) 
that describes the change in
the tunneling amplitude will both flip the qubit (remember that
$\displaystyle{\sigma_z=\rho_z \cos\theta - \rho_x \sin\theta}$) and 
change the occupation of the trap. Thus, it contributes 
to $\Gamma_1$ already in the first order:
\begin{equation}
 \Gamma_1^{(1)} = 2\pi \tilde{t}^2 g(E)\sin^2\theta.
\label{g1}
\end{equation} 
Here $g(\omega)$
is the density of states of excitations. 
The electrostatic interaction term, $\frac{v}{2} \sigma_z
c^{\dagger} c$, does not change the occupation of the localized state,
so it contributes only in the second order. Assuming that
$\displaystyle{(t_0 g_0)^2/E <<g(E)}$, where $g_0$ is the 
density of states in the metal, we can write this contribution as
\begin{equation}
 \Gamma_1^{(2)} =\frac{2\pi}{E^2}v^2[t_0+2\tilde{t}\cos\theta]^2
 g(E)\sin^2\theta 
\label{g2}
\end{equation} 
For {\em Model II},  the contribution (\ref{g2}) does not appear as long as
 all traps are coupled equally to the qubit. Then 
moving one electron from one trap to another will not change the electrostatic
potential. Accordingly for {\em Model II} the $v^2$ should be interpreted
as $\langle v^2\rangle$ averaged over some scatter of $v$.  

Let us calculate $g(\omega)$ for {\em Models I-III}. 
Consider the density of localized state:
\begin{equation}
\nu(\epsilon) = \bar{\nu} + \delta \nu (\epsilon),
\label{dos}
\end{equation}
where $\bar{\nu}$ is the average value of  $\nu(\epsilon)$ (we assume that 
$\bar{\nu}$ is $\epsilon$-independent). The random deviations  $\delta \nu (\epsilon)$ are assumed to be small, $\displaystyle{\delta \nu(\epsilon) \ll \bar{\nu}}$ and only short-range correlated:
\begin{equation}
\langle \delta \nu(\epsilon) \delta \nu(\epsilon') \rangle 
  = A \delta(\epsilon - \epsilon')\;
\label{correlation}
\end{equation}
For the density of states $\langle g(\omega) \rangle$ 
averaged over different realizations of 
the random distribution of trap energies, we have 
\begin{eqnarray}
g(\omega)^{(I)} &=& g_0 \int_{-\omega}^{\omega} d \epsilon \,\nu(\epsilon), \qquad 
   \quad
 \langle g(\omega) \rangle^{(I)}=2 \bar{\nu} g_0 \omega \nonumber \\ 
g(\omega)^{(II)} &=& \int_{0}^{\omega} d \epsilon \,\nu(\epsilon) \nu(\epsilon-\omega),
  \quad \langle g(\omega) \rangle^{(II)}=\bar{\nu}^2 \omega  \nonumber \\
g(\omega)^{(III)} &=& \int_0^\omega d\epsilon\,[\nu(\epsilon)\nu(\omega-\epsilon)
\label{j1}  \\
&+&\nu(-\epsilon)\nu(\epsilon-\omega)],
\quad
\langle g(\omega) \rangle^{(III)}=2 \bar{\nu}^2 \omega.\nonumber \;
\end{eqnarray}
$g_0$ denotes the density of states in the metal and we neglect its 
energy dependence. 

In each of the three models 
$\langle g(\omega) \rangle$ is a linear function of
the frequency $\omega$.

This statement may become incorrect when Coulomb interaction between 
trapped electrons is taken into account. For example in 
{\em Model II} 
the tunneling amplitude depends exponentially on the distance between the 
traps. Only traps which are close in space can exchange charge. For such pairs
the 
Coulomb interaction between the traps 
modifies the density of states to (see Ref. \onlinecite{efros}) 
$\langle g(\omega)\rangle^{(II)}=\bar{\nu}^2 (\epsilon_c^{(II)}+\omega)$.
where $\epsilon_c$ is the energy of the Coulomb interaction between two 
electrons that occupy the two traps. $\epsilon_c^{(II)}$ can be estimated 
as $\epsilon_c^{(II)}=e^2/r_t$, where $r_t$ is the typical distance 
between the traps, i.e. typical tunneling length

As to the {\em Model III} the Coulomb repulsion leads to  
$\langle g(\omega)\rangle^{(III)}=2\bar{\nu}^2 (\omega-\epsilon_c^{(III)})$.
When estimating $\epsilon_c^{(III)}$ we need to take into account the screening
 provided by the superconductor \cite{kozub}: each electron trapped near the 
superconductor creates an image charge and thus forms a dipole moment
of the order of $er_t$. The distance between the two traps is determined by 
the coherence length $\xi$ of the superconductor. Therefore 
$\epsilon_c^{(III)} \sim e^2r_t^2/\xi^3$.

It is informative to compare the Coulomb energies $\epsilon_c^{(II,III)}$
with the superconducting gap $\Delta\sim\hbar v_F/\xi$, where $v_F$ is the 
Fermi velocity \footnote{We considered here the ``clean'' case , when the 
electronic mean free path exceeds $\xi$. Taking disorder into 
account does not change the final conclusion}. Since $e^2/\hbar v_F\approx1$,
we have $\epsilon_c^{(II)}/\Delta\sim\xi/r_t$, and $\epsilon_c^{(III)}/\Delta
\sim (r_t/\xi)^2$. It is natural to assume that $r_t\ll\xi$. Therefore 
$\Delta\ll\epsilon_c^{(II)}$ and  $\Delta\gg\epsilon_c^{(III)}$. 
When considering JQC one is interested in $\omega\sim E_J\ll\Delta$. 
We conclude that $\omega\ll\epsilon_c^{(II)}$ and thus  
  $\langle g(\omega)\rangle^{(II)}=const$. At the same time 
 $\omega\gg\epsilon_c^{(III)}$, $\langle g(\omega)\rangle^{(III)}$ 
is determined by Eq (\ref{j1}) and the ``Andreev fluctuators'' can 
lead to a linear dependence of $\Gamma_1$ on $E_J$.

Returning now to the equations (\ref{g1}) and (\ref{g2}) we see that the 
first order contribution is directly proportional to the density 
of states.
The second order term, because of the $E$ in
the denominator, does not give a linearly increasing relaxation rate
even if the density of states is linear. 

We conclude that to get a linear rate from a linear density of states 
we need some term in the Hamiltonian that gives a contribution to 
first order.

So far we only considered the average relaxation rate, let us now turn 
to the fluctuations. 

Consider the two-point correlator:
\begin{eqnarray}
\langle g(\omega) g(\omega') \rangle_c^{(I)}
  &=& 2 A g_0^2 \min(\omega,\omega'), \nonumber\\ 
\langle g(\omega) g(\omega') \rangle_c^{(II)}
  &=& 2 A \bar{\nu}^2 \min(\omega,\omega')
  +A^2 \omega\delta (\omega-\omega'), 
\label{j2}\\ 
\langle g(\omega) g(\omega') \rangle_c^{(III)}
  &=& 8 A \bar{\nu}^2 \min(\omega,\omega')+
    4 A^2 \omega \delta (\omega-\omega').\nonumber \;
\end{eqnarray}
Note that from Eqs. (\ref{j1}) and (\ref{j2}) it follows that in 
{\em Model I} $g(\omega)$ is a monotonous function of $\omega$,
whereas {\em Models II, III} lead to non-monotonous fluctuations with 
short range correlations.

\begin{figure}[h]
\psfrag{x}{$\hspace{-3mm}$ \large{$\Gamma_1$}}
\psfrag{y}{\large{$\omega$}}
\includegraphics[width=8cm]{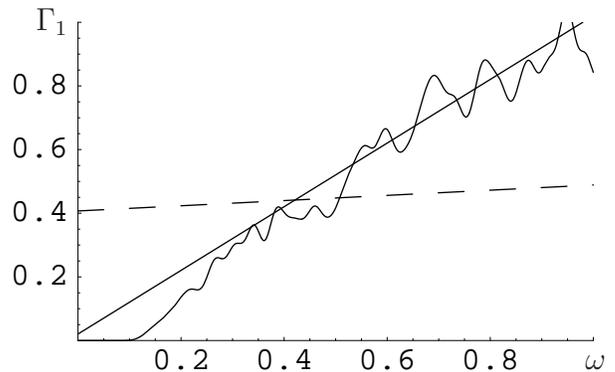}
\caption{The relaxation rate as function of frequency for {\em Model III}. 
The curve shows 
one typical realization of the disorder while the solid line is the 
average over realizations. The dashed line shows $\langle\Gamma_1(\omega)^2
\rangle$.
\label{fig2}}
\end{figure}

Figure \ref{fig2} shows for {\em Model III} at the optimal point
the relaxation rate as function of frequency
for a particular realization of the position in energy of the traps. 
The straight lines represent the result of an ensemble averaging.  If
the fluctuations are rapid one can instead average over suitable
frequency windows for a single sample.

The second order term can also give rise to fluctuations in the 
relaxation rate. Let us focus on {\em Model I} at the optimal point 
($\cos\theta=0$). The main source of fluctuations is the $v^2$-term
and for this 
it follows immediately from Eqs (\ref{g2}) and (\ref{j2}) 
that the correlator is  
\begin{equation}
\langle \Gamma_1^{(2)}(\omega) \Gamma_1^{(2)}(\omega') \rangle_c^{(I)}
  = 2 A g_0^2 \frac{\min(\omega,\omega')}{\omega^2\omega'^2}. 
\end{equation}
Thus, also in {\em Model I} there will be fluctuations, but the peaks will have
a different shape, and the correlations are long range. 
Also, in {\em Models II,III} the amplitude of the oscillations 
increases with increasing 
$E$, whereas in {\em Model I} it decreases since it only has 
contributions to second order. This could be a way to distinguish 
the different models. 
\begin{figure}[h]
\psfrag{y}{$\hspace{-3mm}\vspace{-20mm}$ \Large{$\Gamma_1$}}
\psfrag{x}{\Large{$\omega$}}
\includegraphics[width=8cm]{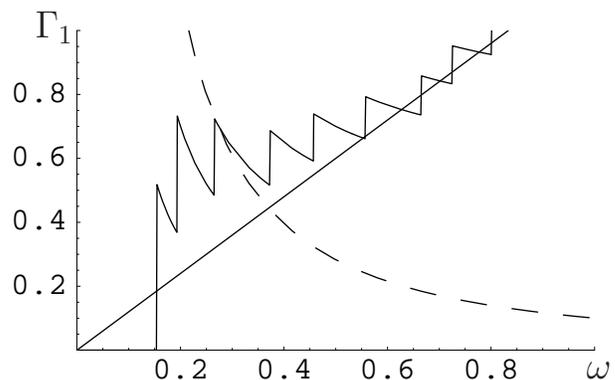}
\caption{The relaxation rate as function of frequency for {\em Model I}. 
The curve shows 
one typical realization of the disorder while the solid line is the 
average over realizations. The dashed line shows $\langle\Gamma_1(\omega)^2
\rangle$.
\label{fig3}}
\end{figure}
For comparison with Figure \ref{fig2} is shown in Figure \ref{fig3}, 
 $\Gamma_1$ for {\em Model I} for the case where $v^2t_0^2/\tilde{t}^2=10$ 
in the units of $E$.

Above we took the correlation of the levels to be a true $\delta$-function. 
In reality, the levels will be broadened by relaxational processes that 
go beyond our models. To take this into account we can use instead for the 
correlator 
some  $\delta$-like function $\delta_\sigma(\epsilon)$ with characteristic 
width $\sigma$. For example, one can think about Gaussian, $\displaystyle
\delta_\sigma^{(G)}=e^{-\epsilon^2/2\sigma^2}/\sqrt{2\pi}\sigma$
or Lorentzian,  $\displaystyle
\delta_\sigma^{(L)}= \sigma/\pi(\epsilon^2+\sigma^2)$.
correlations.
The $\delta$-functions in Eq (\ref{j2}) are then replaced by
the function $\tilde\delta_{\tilde{\sigma}}$ which 
 again is  a $\delta$-like function;
in the Gaussian case
$\tilde\delta_{\tilde{\sigma}}^{(G)}=\delta_{\sqrt{2}\sigma}^{(G)}$ and for 
Lorentzian, 
$\tilde\delta_{\tilde{\sigma}}^{(L)}=\delta_{2\sigma}^{(L)}$.

Note that both phonon and
photon radiation could cause a linear frequency  dependence of the
relaxation rate  in two dimensions due to their linear dispersion.
However we do not believe that they can be responsible for
the observed resonances.
In this case a
peak in the $\Gamma_1$ as function of $E$ follows directly from
a resonance in the density of states $g(E)$.
Let us estimate frequency of such a resonance 
assuming that the resonant structure
in the density of states arise from quantization of phonon levels in
a confined geometry. According to Ref.~\onlinecite{astafiev} one such
resonance was at a frequency of 30 GHz.
The sound velocity is $10^3$ m/s and corresponds to a wavelength of
30 nm. While not impossibly small, this appears to be smaller than the
typical sizes of $>$ 100 nm of the
structures in the samples used.
On the other hand the possibility that there could be a coupling
to a standing photon mode in the experimental cavity looks more likely.
A similar argument but using
the speed of light gives us a wavelength of 1 cm, which is of the right
order of magnitude. Only two samples where measured~\cite{astafiev} with
slightly different resonant frequencies (20 and 30 GHz), but this could be
caused by different position or size of the sample. In view of the fact that
the cavity contains the sample and mount as absorbing material and that no
special care was taken to create a high Q cavity it appears unlikely that
such a sharp resonance line would be created. This could be tested by
introducing some absorbing material into the cavity and see if the
resonant peaks will disappear.
An alternative way to discriminate between a phonon or photon resonance 
peak and one created by a resonant fluctuator 
would be to thermocycle a given sample.
If the resonance is caused
by some fluctuator, the latter probably would be rearranged
by the heating, and thus the peak positions would shift.

In summary, we have argued that dephasing by phonons or photons is unlikely
to explain the experimental results although they can not be ruled out
conclusively.
A more likely explanation is some resonant fluctuator model.
We have discussed three such models, and all of them depend on the effect 
of the state of the qubit affecting the barrier height to reproduce the 
linear dependence of the relaxation rate on $E$. 

We think that  
{\em Model III} (Andreev fluctuators) 
is the most promising for the following reasons.
{\em Models II, III}
allow for rapid, nonmonotonous 
oscillations of the $\Gamma_1$ for large $E$, whereas {\em Model I}
to first order only will show steplike monotonous increase of $\Gamma_1$. 
To second order there are nonmonotonous oscillations also for {\em Model I} 
but they have a different shape. {\em Model II} is less likely than 
{\em Model III} because the Coulomb interaction most likely changes the 
density of states to constant for the relevant range of energies. 

To experimentally determine the coupling constants we suggest the following:
 If one probes the same energy $E$ at different working points
 (by changing both $\delta E_c$ and $E_J$) there should to first order be 
 collapse of the data points if one plots $\Gamma_1/\sin^2\theta$ as a function
of $E$, 
 whereas the terms with $\cos\theta$ in 
 $\Gamma_1^{(2)}$ will cause some deviation. In particular, it seems 
instructive to plot 
$
[\Gamma(E,\theta)/\Gamma(E,\pi/2)-1]/\cos\theta = 
4(v/E)^2(t_0/\tilde{t}+\cos\theta)
$
as function of $\cos\theta$. From this one could extract 
 the ratios $v/E$ and $t_0/\tilde{t}$. 

We proposed thermocycling 
as an experimental check
for the presence of fluctuators
 and introduction of some absorbing material in the 
cavity to rule out photon resonances. 

\acknowledgments
This work was supported by the Norwegian Research Council, grant 153206/V30, 
NSF DMR-0210575, DARPA-QUIST  and
the U. S. Department of Energy Office
of Science through contract No. W-31-109-ENG-38.
We are grateful to O. Astafiev, Yu. A. Pashkin, Y. Nakamura, T. Yamamoto, J. S. Tsai and I. Lerner for discussions.


\begin{thebibliography}{99}
\bibitem{VION02} D. Vion, A. Aassime, A. Cottet, P. Joyez, H. Pothier, C. Urbina, D. Esteve, and M. H. Devoret, {\sl Science} {\bf 296}, 886 (2002).
\bibitem{NAKA99} Y. Nakamura, Yu. A. Pashkin, and J. S. Tsai, {\sl Nature} {\bf 398}, 786 (1999).
\bibitem{YU} Y. Yu, S. Han, X. Chu, S. Chu, and Z. Wang, {\sl Science} {\bf 296}, 889 (2002).
\bibitem{MARTI02} J. M. Martinis, S. Nam, J. Aumentado, and C. Urbina, {\sl Phys. Rev. Lett.} {\bf 89}, 117901, (2002).
\bibitem{CHIORES} I. Chiorescu, Y. Nakamura, C. Harmans, and J. Mooij, {\sl Science} {\bf 299}, 1869 (2003).
\bibitem{PASHK03} Yu. A. Pashkin, T. Yamamoto, O. Astafiev, Y. Nakamura, D. V. Averin, and J. S. Tsai, {\sl Nature} {\bf 421}, 823 (2003).
\bibitem{astafiev} O. Astafiev, Yu. A. Pashkin, Y. Nakamura, T. Yamamoto and J. S. Tsai, \prl {\bf 93}, 267007 (2004).
\bibitem{efros} A. L. Efros and  B. I. Shklovskii, 
in A. L. Efros and M. Pollack (eds.): Electron-Electron 
interactions in Disordered Systems, Elsevier 1985. 
\bibitem{kozub}
V. I. Kozub, A. A. Zuzin, Y. M. Galperin, and V. Vinokur, cond-mat/0411379 (2004).
\end{thebibliography}
\end{document}